\documentclass[10pt,journal]{IEEEtran}
\IEEEoverridecommandlockouts
\usepackage{cite}
\usepackage{amsmath,amssymb,amsfonts}
\usepackage{algorithmic}
\usepackage{graphicx}
\usepackage{textcomp}
\usepackage{booktabs} % For formal tables
\usepackage{url}
\usepackage{courier}
\usepackage{makecell}
\usepackage{epsfig}
\usepackage{multirow}
\usepackage{subfigure}
\usepackage{multicol}
\usepackage{nomencl}
\usepackage{framed}

\def\BibTeX{{\rm B\kern-.05em{\sc i\kern-.025em b}\kern-.08em
    T\kern-.1667em\lower.7ex\hbox{E}\kern-.125emX}}

\newcommand{\Comment}[1]{}
\usepackage[disable]{todonotes}   % Adding prefix % to the above line and remove the prefix % for this line can disable showing the comments

\begin{document}
\title{CoMID: Context-based Multi-Invariant Detection for Monitoring Cyber-Physical Software}

\author{Yi~Qin, Tao~Xie, Chang~Xu, Angello~Astorga, and Jian~Lu
\IEEEcompsocitemizethanks{\IEEEcompsocthanksitem
Y. Qin, C. Xu and J. Lu are with the State Key Laboratory for Novel Software Technology, and the Department of Computer Science and Technology, Nanjing University, Nanjing, China (email: borakirchies@163.com, changxu@nju.edu.cn, lj@nju.edu.cn).
\IEEEcompsocthanksitem T. Xie and A. Astorga are with the Department of Computer Science, University of Illinois at Urbana-Champaign, Urbana, Illinois (email: taoxie@illinois.edu, aastorg2@illinois.edu).}
}
\maketitle

\begin{abstract}
Cyber-physical software continually interacts with its physical environment for adaptation in order to deliver smart services. However, the interactions can be subject to various errors when the software's assumption on its environment no longer holds, thus leading to unexpected misbehavior or even failure. To address this problem, one promising way is to conduct runtime monitoring of \emph{invariants}, so as to prevent cyber-physical software from entering such errors (a.k.a. \emph{abnormal states}). To effectively detect abnormal states, we in this article present an approach, named \underline{Co}ntext-based \underline{M}ulti-\underline{I}nvariant \underline{D}etection (CoMID), which consists of two techniques: \emph{context-based trace grouping} and \emph{multi-invariant detection}. The former infers contexts to distinguish different effective scopes for CoMID's derived invariants, and the latter conducts ensemble evaluation of multiple invariants to detect abnormal states. We experimentally evaluate CoMID on real-world cyber-physical software. The results show that CoMID achieves a 5.7--28.2\% higher true-positive rate and a 6.8--37.6\% lower false-positive rate in detecting abnormal states, as compared with state-of-the-art approaches (i.e., Daikon and ZoomIn). When deployed in field tests, CoMID's runtime monitoring improves the success rate of cyber-physical software in its task executions by 15.3--31.7\%.
\end{abstract}


\listoftodos

\section{Introduction}
\label{sec:introduction}
\IEEEPARstart{C}{yber-physical} software programs (or \emph{cyber-physical programs}) continually interact with their  external physical environments to provide context-aware adaptive functionalities. Examples of such programs  include those running on robot cars~\cite{yang@ase14,car2,car3}, unmanned aerial vehicles (UAVs)~\cite{sebastian,uav2,uav3}, and humanoid robots~\cite{nao,robot2,robot3}. Cyber-physical programs keep sensing environmental changes, making decisions based on their pre-programmed logics, and then taking physical actions to cope with the sensed changes. The three steps, namely, sensing, decision-making, and action-taking, form an interaction \emph{loop} between a cyber-physical program and its running environment. Each pass of such an interaction loop is referred to as an \emph{iteration}.

To improve the productivity and cope with infinite kinds of environmental dynamics, software developers often hold certain assumptions on typical scenarios, where their cyber-physical programs are supposed to run. For example, a robot controlled by a cyber-physical program walks in an indoor environment, where the floor is supposed to be firm but not slippery, and the space is supposed not to contain any fasting-moving obstacle. However, it is challenging for developers to precisely specify what can be considered as ``not firm'' or ``slippery''. Besides, it is also challenging for users to determine whether a specific scenario meets such vague assumptions, and when the assumptions no longer hold (e.g., when the scenario's humidity has largely reduced its floor's friction force). As such, a cyber-physical program is easily subject to runtime errors in its deployment~\cite{kulkarni@tse10,sama@tse10,xu@jss12,betty@seams12,roadmap}, and then causes misbehavior or even failure (e.g., a robot falling down and making itself damaged). Therefore, there is a strong need for preventing cyber-physical programs from entering such errors, which indicate the violation of their implicit assumptions on running environments.

One promising way is to conduct runtime monitoring of pre-specified invariants, which represent the properties that have to be satisfied during executions, to check whether a cyber-physical program's execution is safe. Being \emph{safe} indicates that the program's execution will not lead to a failure, if no intervention is taken, but just following the logics in the program.  However, specifying effective invariants is challenging. For example, one may specify invariants as the negation of failure conditions, e.g., not crashing of a UAV or falling down of a humanoid robot. However, such invariants are not that useful, since when they are violated (i.e., corresponding failure conditions are evaluated to be true), it is already too late for a concerned program not to fail. Another alternative is to specify invariants for latent erroneous states (a.k.a. \emph{abnormal states}). Then one is potentially able to predict future failures, and prevent a concerned program from taking originally-planned actions, which would otherwise have caused failures. For example, if a robot finds its program execution violating the invariants representing safe executions, it can decide to stop further exploring the current scenario and plan another path to its destination. This resolution action can help it avoid unexpected danger in the original scenario.

There are two major ways of specifying invariants for detecting abnormal states: using \emph{manually specified properties} or using \emph{automatically generated invariants}. For the former, developers need domain knowledge to understand what can constitute abnormal states, and derive corresponding properties. This manual process is challenging, especially when a cyber-physical program and its running environment are non-trivial~\cite{sebastian}. On the other hand, automated invariant generation~\cite{daikon,inv1,inv2,inv3} provides a promising alternative. Despite varying in details, these approaches follow a general process~\cite{nguyen@fse13} as follows. When a subject program is running, they collect its execution trace in terms of program states (e.g., variable values) at  program locations of interest (e.g., entry and exit points of each executed method). Then from a set of such collected safe traces (i.e., no led failure), the approaches derive invariants for different program locations based on predefined templates. These invariants can then be used with runtime monitoring to predict the program's future executions to be safe (i.e., \emph{passing}, for no invariant violation) or not (i.e., \emph{failing}, for any invariant violation). Here, passing implies that the program runs safely with its assumptions on the environment holding, and failing implies that the program could soon fail since its assumptions on the environment no longer hold now.

However, using automatically generated invariants for runtime monitoring is still challenging. One major problem is how to balance between \emph{general} and \emph{specific} invariants. If an invariant for a program location is too general, using it for runtime monitoring can miss the detection of abnormal states, resulting in false negatives. For example, relaxing invariants to cater for various firm floors can accidently include firm but slippery floors, breaking the robot program's assumptions on its running environment. On the other hand, if an invariant is too specific, using it for runtime monitoring can detect many ``abnormal'' states even in safe executions, resulting in false positives. For example, restricting invariants to specific firm floors (e.g., in brick or wood material) can cause false warnings when the robot walks on other firm but not slippery floors, where the program's execution is still safe.

Even worse, this balancing problem can be further exacerbated by two characteristics of cyber-physical programs:

\textbf{Iterative execution.} Cyber-physical programs are featured by repeated iterations of a sensing, decision-making, and action-taking loop. Then a program location for which an invariant is generated can be executed multiple times during \emph{multiple iterations} for dealing with \emph{different contexts} (i.e., various situations in handling environmental dynamics). During these different iterations, a program's definition of \emph{safe} behavior with respect to each context varies across the iterations. Overlooking these contexts, generated invariants would be overgeneralized, such that the detection of abnormal states can be missed. On the other hand, generating invariants by sticking to any specific context would also make the invariants overly fragile to other contexts of safe executions, causing false warnings.

\textbf{Uncertain interaction.} Cyber-physical programs could also face massive false alarms due to \emph{uncertainty}~\cite{esfahani@fse11}, when they use automatically generated invariants to detect abnormal states. For example, even if one places a robot at the same position across different iterations, its sensors can possibly report different values for its position due to uncertainty (as an inherent nature of sensing). These different input values are then propagated to a program location of interest for deriving invariants, causing this location to own variable values different from those in other safe executions also from the same position. Then, overlooking the impact of such uncertainty, runtime monitoring with generated invariants can easily report false alarms: invariant violation is actually caused by inaccurate sensing, not due to a program's assumptions not holding on its environment.

To address these challenges, we in this article present an approach, named \underline{Co}ntext-based \underline{M}ultiple \underline{I}nvariants \underline{D}etection (\emph{CoMID}), to automatically generating invariants and using them for effective detection of abnormal states for cyber-physical programs. CoMID addresses the preceding challenges by its two techniques, namely, \emph{context-based trace grouping} and \emph{multi-invariant detection}:

\textbf{Context-based trace grouping.} The first technique groups execution traces collected from different iterations in a cyber-physical program's execution according to their contexts. Then the execution traces in each group include only those from the iterations that share the same program and environmental contexts. Here, \emph{program context} refers to a program's statements executed during one iteration, and \emph{environmental context} refers to the values of environmental attributes as sensed by the program during the iteration. This technique conducts execution trace grouping by clustering, based on the similarities of corresponding contexts between each pair of execution traces. Then, for each group the technique generates invariants based only on the execution traces in that group. Since the execution traces in a group share a common program context and environmental context, the two contexts together specify the effective scope for the invariants generated for this group. We name it the group's generated \emph{invariants' context}, which is used for runtime monitoring: an invariant is valid to check (i.e., within the effective scope), only when its context is close enough to that of a program's current iteration. In this case, the iteration is named \emph{context-sharing iteration}. By doing so, CoMID aims to both avoid missing the detection of abnormal states and avoid reporting false alarms due to checking an invariant in a non-context-sharing iteration.

\textbf{Multi-invariant detection.} The second technique addresses the robustness problem for invariants when their relied execution traces contain noisy values due to uncertainty. Instead of generating a single invariant from \emph{all} execution traces in a group, this technique generates multiple ones, based on \emph{different subsets} sampled from the execution traces in the group. Then it uses an estimation function to decide the detection of abnormal states based on multi-invariant evaluation results. The function measures the ratio of violated invariants against all invariants with respect to their corresponding groups, and then takes the uncertainty in program-environment interactions into consideration, to decide whether the invariant violation indicates the detection of abnormal states or is simply caused by uncertainty. This idea has been inspired by ensemble learning~\cite{em}, which uses multiple models to improve the prediction performance, as compared with the conventional prediction based on one constituent model alone.

We experimentally evaluate our CoMID approach on three real-world cyber-physical programs: a 4-rotor unmanned aerial vehicle (4-UAV) \cite{boshi}, a 6-rotor unmanned aerial vehicle (6-UAV), and a NAO humanoid robot \cite{nao}. We compare CoMID with two state-of-the-art approaches on automated invariant generation: Daikon~\cite{daikon} and ZoomIn~\cite{zoom}. The experimental results show CoMID's effectiveness: it achieves a 5.7--28.2\% higher true-positive rate and a 6.8--37.6\% lower false-positive rate in detecting abnormal states for the three programs' executions; when deployed for runtime monitoring to prevent unexpected failures, CoMID improves the success rate of the three programs by 15.3--31.7\% in their task executions.

In summary, this article makes the following contributions:

\hangafter 1
\hangindent 1.5em
$\bullet$ The CoMID approach to automatically generating invariants and detecting abnormal states for cyber-physical programs' executions.

\hangafter 1
\hangindent 1.5em
$\bullet$ The context-based trace grouping technique to refine invariant generation with respect to different contexts.

\hangafter 1
\hangindent 1.5em
$\bullet$ The multi-invariant detection technique to address the impact of uncertainty in program-environment interactions on invariant-based runtime monitoring.

\hangafter 1
\hangindent 1.5em
$\bullet$ An experimental evaluation with real-world cyber-physical programs and comparison of CoMID with state-of-the-art invariant generation approaches.

The remainder of this article is organized as follows. Section~\ref{sec:example} uses a motivating example to illustrate our target problem and its challenges. Section~\ref{sec:method} gives an overview of our CoMID approach and then elaborates on its two techniques. Section~\ref{sec:exp} presents our evaluation of CoMID with three real-world cyber-physical programs and compares it with existing approaches. Section~\ref{sec:rela} discusses the related work in recent years, and finally Section~\ref{sec:end} concludes this article and discusses future work.

\section{Motivating Example}
\label{sec:example}
\begin{figure}
\subfigure[Walking on a wood floor]{
\label{fig:example2} %% label for first subfigure
\includegraphics[width=0.23\textwidth]{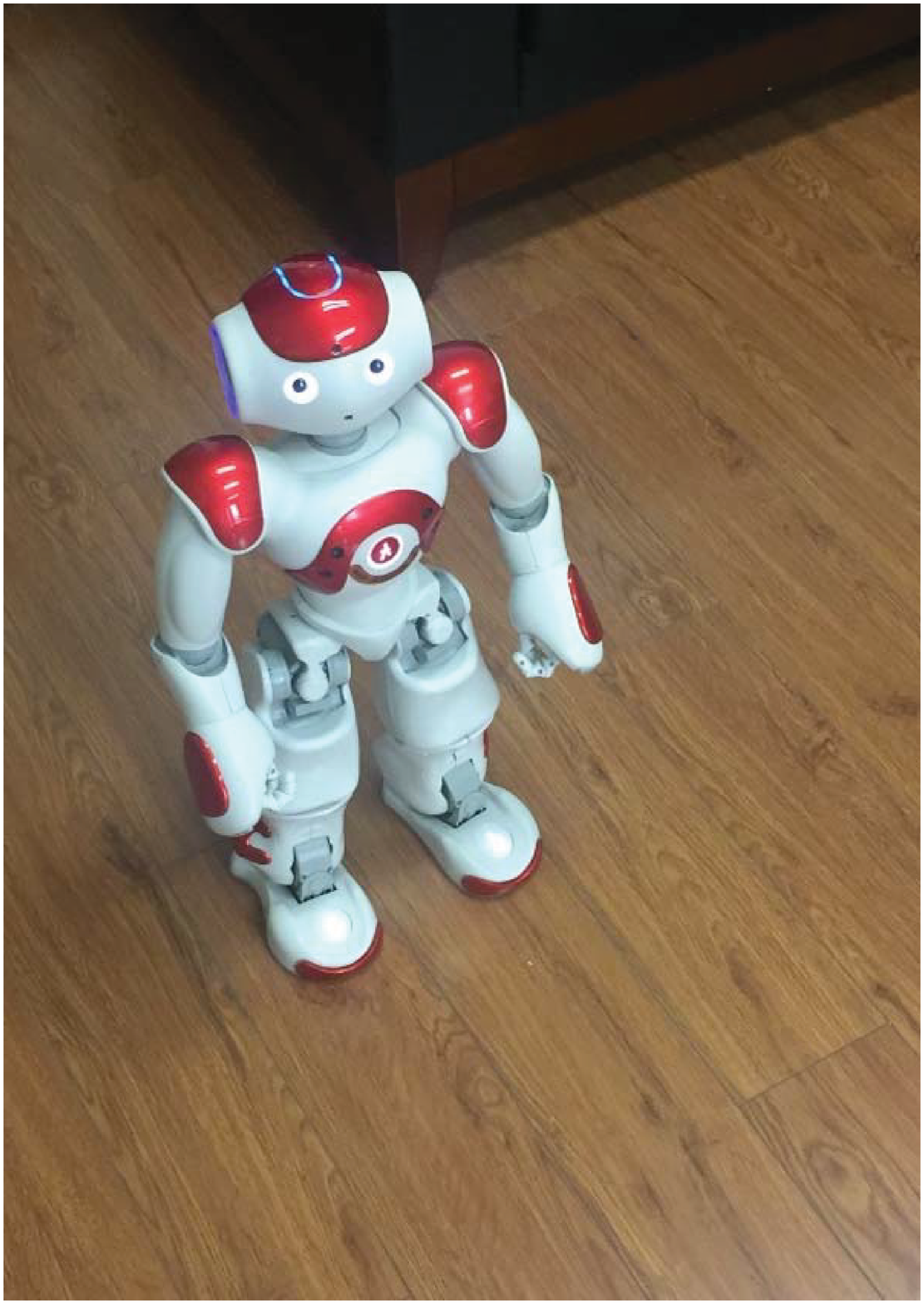}}
\subfigure[Walking on a brick floor]{
\label{fig:example2} %% label for first subfigure
\includegraphics[width=0.23\textwidth]{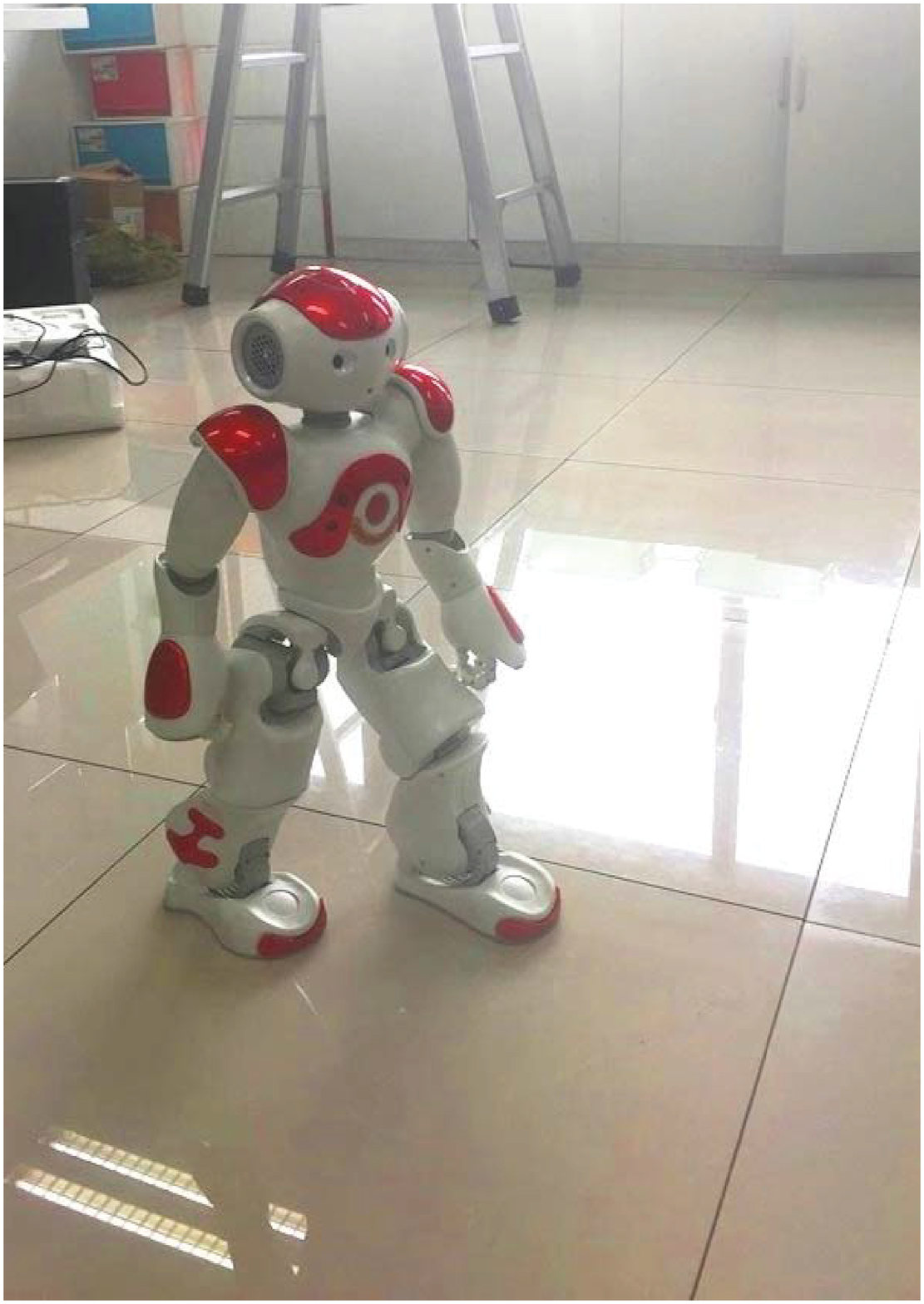}}
\caption{A NAO robot controlled by a cyber-physical program}
\label{fig:example}
\end{figure}

In this section, we use a motivating example to illustrate our target problem and its challenges. Consider our aforementioned NAO humanoid robot controlled by a cyber-physical program $P$. Suppose that the robot is exploring an indoor area, as illustrated in Fig.~\ref{fig:example}. For the sake of quality and productivity, developers can hold implicit assumptions on the scenarios where the robot is supposed to walk, e.g., a room with a firm and not slippery floor. Then, developers proceed to design corresponding exploration strategies for the robot, e.g., walking slowly and balancing by raising its arms with certain angles. These strategies are for ensuring the robot to walk safely on a floor made of several common materials, e.g., wood, as shown in Fig.~\ref{fig:example}-a, and brick, as shown in Fig.~\ref{fig:example}-b. In the following, we analyze what challenges the runtime monitoring with invariants can encounter, in order to prevent the robot from entering abnormal states.

Program $P$ uses readings of two pressure sensors installed on the robot's two feet to measure whether the robot has leaned toward left or right and decide whether it has to balance the robot in its walking. The measurement is conducted by calculating the difference between the two sensors' readings, $pre_{left}$ and $pre_{right}$. $P$ then decides one of the robot's arms according to which direction the robot is leaning toward, and calculates the height the decided arm should be raised to. Suppose that variable $angle$ in $P$ controls the height value, and then it becomes a key factor that decides whether the robot can properly balance itself in walking. Developers can design various logics to calculate the $angle$ value, but they more or less depend on the material comprising the floor.

One outstanding challenge is that developers can hardly specify proper $angle$ values. They typically follow a trial-and-error process to calculate plausible $angle$ values. If lucky enough, they can design the calculation logics that seemingly work for several types of floor material. Even so, users may still not be able to decide whether a specific scenario is safe for the robot to walk into (i.e., whether the calculation logics still work), or when a previously safe scenario becomes no longer safe (e.g., when the scenario gradually evolves). As mentioned earlier, this is the right place for runtime monitoring with invariants plays an important role. In the following, we explain how to generate invariants for the $angle$ variable and use them to decide whether $P$'s execution is safe for the current scenario.

Most existing invariant generation approaches work similarly. Now we generate an invariant for variable $angle$ at the entry point of method {\fontfamily{\ttdefault}\selectfont motion.}{\fontfamily{\ttdefault}\selectfont angleMove} {\fontfamily{\ttdefault}\selectfont (names,} {\fontfamily{\ttdefault}\selectfont angle,} {\fontfamily{\ttdefault}\selectfont timeLists)}, which is the key method for deciding how to raise an arm for balancing the robot. We first collect several safe execution traces (e.g., $tr_1$, $tr_2$, and $tr_3$) of program $P$, in which $angle$'s corresponding variable-value pairs are $tr_1$: $\{angle = 48\}$, $tr_2$: $\{angle = 52\}$, and $tr_3$: $\{angle = 55\}$. Following a predefined template (e.g., $varX \leq C$), we can derive an invariant like ``$angle \leq 55$'', satisfying all the three traces. This invariant suggests that proper $angle$ values at this program location should not exceed 55. Then later when $P$ controls the robot and finds its collected $angle$ value at the same program location to be 60, the runtime monitoring could decide that $P$'s execution is not safe. Technically, it reports that the current execution enters an abnormal state, i.e., classified as \emph{failing}.

However, as mentioned earlier, invariant generation has to balance between general and specific invariants. The preceding invariant ``$angle \leq 55$'' has relaxed its condition on proper values for the $angle$ variable to cater for all the three execution traces, although these values could be from different scenarios. Then using this invariant can potentially misclassify an unsafe execution with an $angle$ value of 53 for the scenario experienced in $tr_1$ as passing. On the other hand, if one derives the invariant from two execution traces, $tr_1$ and $tr_2$, only (e.g., ``$angle \leq 52$''), but checks it against the execution of $tr_3$ from another scenario. Then the runtime monitoring can be too strict and would misclassify that execution as \emph{failing}.

\begin{figure*}
\centering \includegraphics[width= 0.9\textwidth]{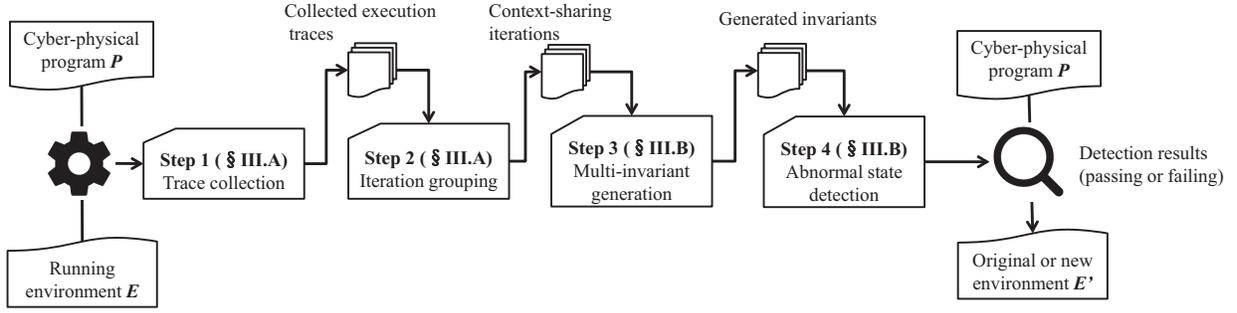}
\centering \caption{CoMID's workflow}
\label{fig:overview}
\end{figure*}

The nature of cyber-physical programs exacerbates the invariant-balancing problem. For example, a cyber-physical program can encounter multiple iterations, and not all iterations share the same context. Suppose that a robot is walking in a scenario connected with different types of floor material (e.g., wood, brick, and others) and placed with different types of obstacle (e.g., high, low, round, and others). Such a scenario implies different \emph{environmental contexts}. Even on the same floor, the robot may take different strategies to handle different obstacle situations. Such variety of strategies implies different \emph{program contexts}. Without distinguishing these contexts, invariant generation can be easily over-generalized (e.g., deriving invariants to cater for all executions traces), and invariant violation can also be easily over-triggered (e.g., checking invariants in a context different from the context from which the invariants are derived).

A cyber-physical program's uncertain interactions with its environment similarly worsen the invariant-balancing problem. Inaccurate sensing makes derived invariants imprecise due to random noises in sensor readings, and the imprecision can cause both false-warning and missing-warning problems. A naive way is to relax the condition in such an invariant by allowing some extent of error, e.g., a delta of $\pm 5$ added to proper values for the $angle$ variable. However, this way is quite ad hoc, and can also easily aggravate the false-warning and missing-warning problems.

These limitations of existing approaches on automated invariant generation motivate us to develop our CoMID approach, particularly focused on the invariant generation and runtime monitoring for cyber-physical programs. CoMID aims to distinguish different contexts for effective invariant generation and address the impact of uncertainty for effective runtime monitoring with generated invariants. We elaborate on our CoMID's methodology in the next section.

\section{Context-based Multi-Invariant Detection}
\label{sec:method}
%\subsection{Overview}
%\label{sec:overall}

The input of our CoMID approach is a cyber-physical program $P$ and its running environment $E$ (conceptually). We consider $E$ as a black-box program without having to know how $E$ exactly works. Due to the interactions between $P$ and $E$, $P$'s output $O_P$ becomes $E$'s input $I_E$, and $P$'s input $I_P$ for its next iteration comes from $E$'s output $O_E$ based on $E$'s internal logics (e.g., physical laws and domain rules). For invariant generation purposes, we assume the availability of a set of failure conditions (e.g., crashing of a UAV or falling down of a humanoid robot) for deciding whether a cyber-physical program's execution has already failed, as existing work does.

CoMID works in four steps: (1) it first executes program $P$ in environment $E$ to collect safe execution traces, i.e., no failure condition triggered (Step 1: \emph{trace collection}); (2) it then groups iterations from the collected execution traces into multiple sets of \emph{context-sharing iterations}, based on their program and environmental contexts (Step 2: \emph{iteration grouping}); (3) after that, it generates multiple invariants for each group (Step 3: \emph{multi-invariant generation}); (4) finally, it uses the generated invariants to detect abnormal states for program $P$'s future executions (Step 4: \emph{abnormal state detection}). Fig.~\ref{fig:overview} illustrates CoMID's workflow.

In the first two steps, besides collecting traditional artifacts (e.g., arguments and return values for each executed method), CoMID also analyzes program and environmental contexts for each iteration. Regarding the program context, CoMID records what statements are executed in an iteration. Regarding the environmental context, CoMID records attribute values associated with environment $E$. CoMID recognizes $P$'s system calls related to environmental sensing, and uses these calls to record attribute values at the beginning of each iteration. CoMID uses the program context to distinguish an iteration's specific strategy in handling external situations, and uses the environmental context to distinguish different situations $P$ is facing in a specific iteration.

In the last two steps, CoMID generates and checks multi-invariants to address the impact of uncertainty on deciding whether a specific invariant violation is a convincing indication that the current execution is no longer safe. CoMID leverages previous work (e.g., Daikon~\cite{daikon}) for invariant derivation by feeding different sets of sampled iterations.

In the following, we elaborate on CoMID's details.

\subsection{Context-based Trace Grouping (Steps 1 and 2)}
\label{sec:information}
\textbf{Trace collection.} In the first step, CoMID executes program $P$ multiple times in environment $E$, and collects $P$'s trace information from its executions that do not trigger any failure condition. For each collected execution trace, CoMID determines the boundary (i.e., beginning and ending) of each iteration in the trace, by recognizing input points (i.e., system calls related to environmental sensing, e.g., for acquiring pressure sensing values on the NAO robot's feet) and output points (i.e., system calls related to physical actions, e.g., raising the NAO robot's arms) in $P$. Following the sensing, decision-making, and action-taking loop, an iteration starts from the first program location where $P$ takes its sensing input from $E$, and ends before the location where it finishes its physical actions in this round and is about to sense the input for the next round. For saving the cost, CoMID records values of program variables only at entry and exit points of the methods executed in each iteration.

To distinguish different iterations, CoMID also records program and environmental contexts for each iteration. For the program context, CoMID records the statements executed in each iteration through program instrumentation. For the environmental context, CoMID records values of environmental attributes using their involved system calls at the beginning of each iteration (i.e., once CoMID recognizes a new iteration).

Formally, we use \emph{segment} to represent the collected information for each iteration in program $P$'s execution. A segment abstracts $P$'s execution state during an iteration. We use $sg^i$ to represent $P$'s state for its $i$-th iteration: $sg^{i}$ = ($P_{cxt}$, $E_{cxt}$, $M_1$, $M_2$, ..., $M_j$), where

\begin{enumerate}
  \item $P_{cxt}$ represents the $i$-th iteration's program context, which is a set of identities (ids) of statements executed in the iteration.
  \item $E_{cxt}$ represents the $i$-th iteration's environmental context, which is a set of name-value pairs for sensing variables in $P$.
  \item $M_1$, $M_2$, ..., $M_j$ represent a sequence of methods executed in the $i$-th iteration, each of which contains a method's name, arguments, and return value.
\end{enumerate}

\textbf{Iteration grouping.} In the second step, CoMID groups iterations (segments) from collected execution traces, so that each group contains only \emph{context-sharing} ones. Here, contexts refer to program and environmental contexts recorded in the first step.

CoMID analyzes environmental contexts $E_{cxt}$ recorded in segments to discover common patterns shared by iterations. It builds a set of all environmental contexts $ENV\_CONTEXT$, and conducts the $k$-means clustering algorithm~\cite{kmeans} to form different clusters. For the performance consideration, CoMID considers only environmental attributes of numeric types in the clustering. It uses a normalized Euclidean metric to measure the distance between each pair of environmental contexts. Given two environmental contexts $E_{cxt}\_A$ $(a\_A_1, a\_A_2, ..., a\_A_n)$ and $E_{cxt}\_B$ $(a\_B_1, a\_B_2, ..., a\_B_n)$, their distance $dis(E_{cxt}\_A, E_{cxt}\_B)$ is calculated as
\begin{center}
$dis(E_{cxt}\_A, E_{cxt}\_B) = \sum_{i = 1}^{n} \sqrt{\frac{(a\_A_i - a\_B_i)^{2}}{s_{i}^{2}}}$,
\end{center}
where $s_i^2$ is the variance of all values of $E_{cxt}$'s $i$-th attributes in the $ENV\_CONTEXT$ set.

The $k$-means clustering algorithm~\cite{kmeans} requires setting a suitable value for parameter $k$, which decides the maximal size of each formed cluster of environmental contexts. Generally, a small $k$ value can make derived clusters more specific, but it could also increase noises in later classification~\cite{knn2}. Therefore, we choose the grid search~\cite{grid}, a traditional way of conducting parameter optimization in machine learning algorithms, to decide the most suitable value for parameter $k$. Intuitively, the grid search conducts cross-validation on a set of candidate values for the parameter to be optimized, and selects the one with the best performance.

We initially use 30 candidate values for parameter $k$, from $1\%$ of the total number of collected environmental contexts to $30\%$, increasing with a pace of $1\%$. Then we conduct 10-fold cross-validation to decide the most suitable $k$ value. We randomly divide the $ENV\_CONTEXT$ set into ten disjoint subsets of the same size. Nine subsets are merged for training (i.e., \emph{training set}) and the remaining one is for validation (i.e., \emph{testing set}). For each candidate $k$ value, we conduct its corresponding clustering on the training set, resulting in multiple clusters of environmental contexts. With respect to these clusters, the environmental contexts from the testing set are then classified into them. Accordingly, we calculate an average deviation value to measure the performance associated with the specific $k$ value. Let an environmental context from the testing set be $E_{cxt}\_T$, and its classified cluster be $C$ ($E_{cxt}\_1$, $E_{cxt}\_2$, ..., $E_{cxt}\_j$). Then context $E_{cxt}\_T$'s deviation value $div(E_{cxt}\_T)$ is calculated as
\begin{center}
$div(E_{cxt}\_T) =  \frac{1}{j}\sum_{i = 1}^{j} dis(E_{cxt}\_T, E_{cxt}\_i)$.
\end{center}

The \emph{average deviation value} for $k$ is the averaged deviation values of all environmental contexts from the testing set. One would expect this value to be minimized, and thus CoMID selects the $k$ value with the smallest average deviation value after comparing all candidate values. In our field tests of the NAO robot and UAV subjects used later in our evaluation (Section~\ref{sec:exp}), we observe that the selected $k$ value ranges from $17\%$ to $22\%$ of the total number of collected environmental contexts with their corresponding performance being similar. Therefore, we select $20\%$ of the total number as the $k$ value used in CoMID to simplify its implementation and evaluation.

With the $k$ value set for the $k$-means clustering, CoMID derives initial clusters for collected environmental contexts, and their owning segments are also clustered accordingly. Then CoMID refines these initial clusters of segments based on their program contexts, by measuring the similarity of program contexts between segments in each cluster. CoMID uses the Jaccard similarity index~\cite{jaccard} to calculate the Degree of Similarity (DoS) value between each pair of program contexts. Let $P_{cxt}\_{sg}$ be segment $sg$'s program context (i.e., a set of statement ids). Then for two given segments $sg_A$ and $sg_B$, the DoS value between their program contexts $DoS(P_{cxt}\_{sg_A}, P_{cxt}\_{sg_B})$ is calculated as
\begin{center}
$DoS(P_{cxt}\_{sg_A}, P_{cxt}\_{sg_B})  =  \frac{|P_{cxt}\_{sg_A} \cap P_{cxt}\_{sg_B}|}{|P_{cxt}\_{sg_A} \cup P_{cxt}\_{sg_B}|}$.
\end{center}

Then the DoS value between a pair of program contexts ranges from 0 to 1. CoMID considers two segments to have the \emph{same} program context if the DoS value of their program contexts is no less than 0.8. This reference value is set by following existing work~\cite{zoom}. Nevertheless, we also study the impact of different DoS threshold values on CoMID's performance in our later evaluation (Section~\ref{sec:exp}).

Based on this similarity measurement on program contexts, CoMID refines the initial clusters of segments. If two segments in one cluster have the same program context, they are still together in that cluster. Otherwise, they are separated into two clusters. This separation process iterates until no cluster can be refined. Then the final result is a set of groups, each of which contains only segments with the same environmental and program contexts. We also say that each group contains \emph{context-sharing iterations}.

\begin{figure}
\centering \includegraphics[width= 0.5\textwidth]{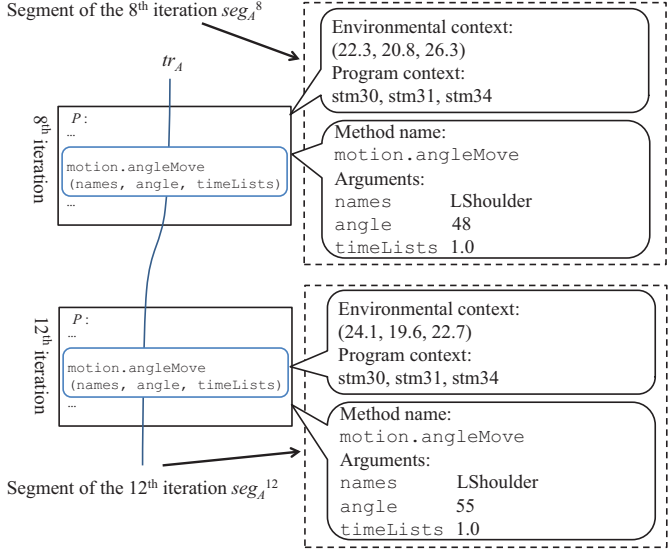}
\caption{Illustration of Step 1: Trace collection}
\label{fig:step1}
\end{figure}

\begin{figure}
\centering
\subfigure[Deriving clusters by environmental context]{
\label{fig:2-1}
\includegraphics[width= 0.45\textwidth]{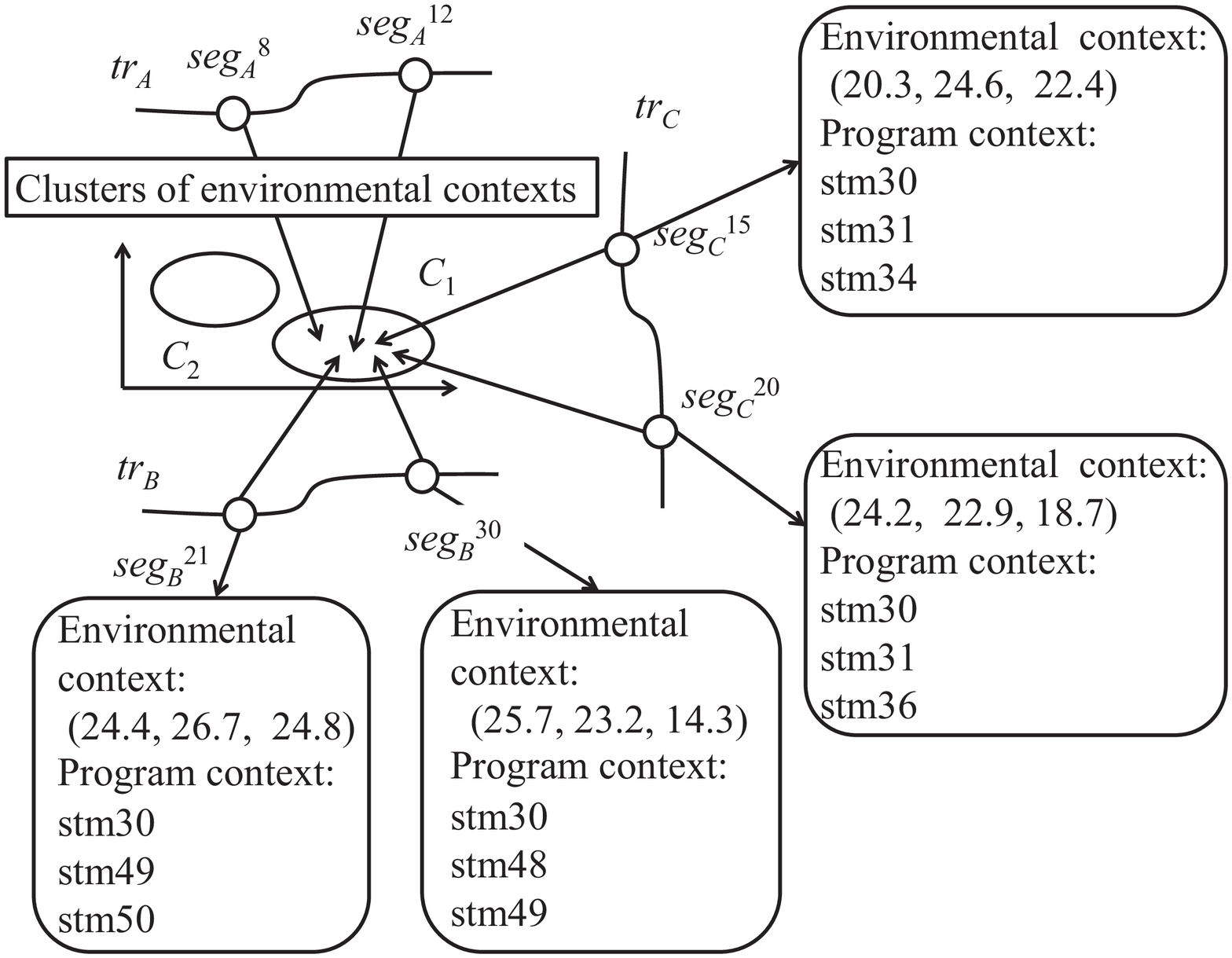}}
\subfigure[Refining clusters by program context to form final groups]{
\label{fig:2-2}
\includegraphics[width= 0.45\textwidth]{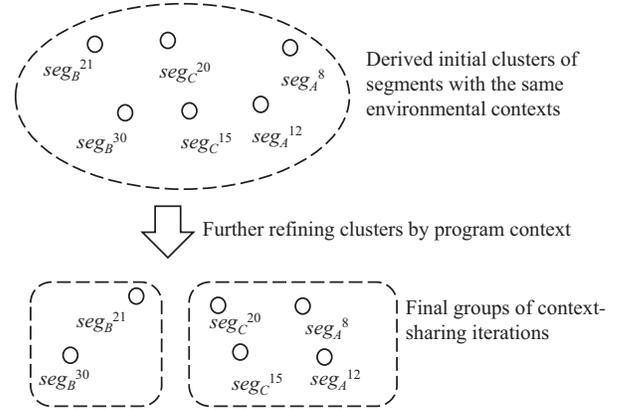}}
\caption{Illustration of Step 2: Iteration grouping}
\label{fig:step2}
\end{figure}

\textbf{Example.} Consider in our robot example method {\fontfamily{\ttdefault}\selectfont motion.}{\fontfamily{\ttdefault}\selectfont angleMove}{\fontfamily{\ttdefault}\selectfont(names,} {\fontfamily{\ttdefault}\selectfont angle,} {\fontfamily{\ttdefault}\selectfont timeLists)}. Fig.~\ref{fig:step1} illustrates CoMID's recorded information for the 8th and 12th iterations in an execution trace $tr_A$. The segment representing the 8th iteration, denoted as $seg_A^8$, is shown in the upper dashed box, and the segment representing the 12th iteration is shown in the lower box. For each segment, its upper block lists the concerned iteration's environmental and program contexts, respectively, and its lower block lists the information for methods executed in this iteration (here we show one method for illustration). We use a tuple, e.g., (22.3, 20.8, 26.3), to represent the values of sensed environmental attributes, e.g., for the pressure on the robot's left foot, that on the right foot, and the robot's distance to its front-facing obstacle, respectively. We use ``stm'' followed by a number, e.g., ``stm30'', to represent the id of a statement executed in the concerned iteration.

Fig.~\ref{fig:step2} illustrates how the iterations in three execution traces ($tr_A$, $tr_B$, and $tr_C$) are grouped according their environmental and program contexts. CoMID first derives initial clusters (Fig.~\ref{fig:step2}-a) according to environmental contexts of the iterations, and cluster $C_1$ includes six iterations ($seg_A^8$ and $seg_A^{12}$ from $tr_A$, $seg_B^{21}$ and $seg_B^{30}$ from $tr_B$, and $seg_C^{15}$ and $seg_C^{20}$ from $tr_C$). We show only their environmental and program contexts for illustration. CoMID then calculates DoS values for program contexts of the six iterations, and refines the $C_1$ cluster into two final groups (Fig.~\ref{fig:step2}-b). One larger group contains four iterations ($seg_A^8$, $seg_A^{12}$, $seg_C^{15}$, and $seg_C^{20}$) from execution traces $tr_A$ and $tr_C$, and the other smaller group contains two iterations ($seg_B^{21}$ and $seg_B^{30}$) from trace $tr_B$. Such refinement result is due to their DoS calculations, e.g., \emph{DoS}($seg_A^8$, $seg_C^{15}$) = 1.0, \emph{DoS}($seg_A^8$, $seg_B^{21}$) = 0.2, and \emph{DoS}($seg_B^{21}$, $seg_C^{15}$) = 0.2, and so on.

\subsection{Multi-Invariant Detection (Steps 3 and 4)}
\label{sec:invariant}
\textbf{Multi-invariant generation.} After context-based trace grouping, CoMID obtains multiple groups of context-sharing iterations in terms of segments. CoMID feeds the segments in each group to the Daikon~\cite{daikon} engine for deriving invariants specific to this group.

As mentioned earlier, CoMID needs to address the impact of uncertainty on invariant generation, so as to suppress the negative consequences of inaccurate sensing values. To do so, CoMID uses different subsets from each group of segments for deriving invariants, which are later used for collective checking in the runtime monitoring against uncertainty. Generally, one can freely decide the number of such subsets, and CoMID chooses four (i.e., 20\%, 40\%, 60\%, and 80\% of the total number of segments in a group) for avoiding high computational and monitoring overheads.

Then, besides the one invariant (i.e., \emph{principal invariant}) for the universal set (i.e., a whole group of segments), CoMID generates four invariants for the four subsets, respectively. These five invariants are named as an invariant \emph{family}, with respect to each supported invariant template and each executed method requiring invariant generation in the group. Since each invariant family is associated with a specific group of context-sharing iterations, the group's contexts are also referred as the \emph{invariant family's context}. An invariant family's context specifies the situations under which the invariants in the family are suitable for checking, thus deciding abnormal states for concerned programs.

\textbf{Abnormal-state detection.} Now CoMID has generated a set of invariant families for runtime monitoring of each program location of interest. Different from existing work, CoMID chooses to check only those invariant families whose contexts are the \emph{same} as that of the current iteration in a program's execution. Here, ``same'' is decided by the comparisons of both program and environmental contexts: (1) the DoS value between a pair of program contexts no less than 0.8 (Section III.A), and (2) the environmental context of the current iteration is classified into the same cluster as that of the considered invariant family.

After selecting suitable invariant families for checking, CoMID then needs to decide whether an invariant violation in the runtime monitoring is simply caused by uncertainty or indicates the detection of a real abnormal state. CoMID uses an \emph{estimation function} to ensemble the evaluation results of invariant checking across multiple iterations, in order to suppress the impact of uncertainty on the decision. The design of the estimation function is based on two \emph{intuitions}:
\begin{enumerate}
  \item The possibility that an invariant violation or satisfaction is caused by uncertainty relates to the number of segments that have been used for deriving the invariant under checking.
  \item The impact of uncertainty on invariant checking can be suppressed by examining checking results across multiple consecutive iterations.
\end{enumerate}

Based on these two intuitions, the estimation function assigns a weight to each invariant violation or satisfaction. The weight assignment is designed as follows:
\begin{enumerate}
  \item For a violated invariant $inv_1$, the more segments are used for deriving it, the less possibility that $inv_1$'s violation is caused by uncertainty, since $inv_1$ is inclined to be general.
  \item For a satisfied invariant $inv_2$, the more segments are used for deriving it, the less possibility that $inv_2$'s satisfaction indicates the current execution to be passing, since satisfying a general invariant is natural.
\end{enumerate}

Recall that CoMID makes five subsets for each group of segments (from 20\% to 100\% of the total size, with a pace of 20\%), and generates invariants with respect to each of these subsets. Then given a subset of segments and its associated size ratio $p$ (i.e., 20\%, 40\%, ..., or 100\%), CoMID sets the weight assigned for the \emph{violation} of one invariant generated from this subset to be $p$, and that for the \emph{satisfaction} to be $-(1-p)$. Such a weight value intuitively models the likelihood whether an execution is failing or passing: a positive value suggests failing, while a negative value suggests passing, and its absolute value indicates the confidence.

Formally, consider an invariant family  $INV$ = \{$inv_i$\}, $1 \leq i \leq k$. Let the invariant-checking result for $inv_i$ at iteration $j$ be $r_{i}^{j}$, where $1$ denotes invariant satisfaction and $-1$ denotes violation. Let the size ratio associated with invariant $inv_i$ be $p_i$ (from its corresponding segment subset). Then the estimation function returns for $INV$ at iteration $j$ as follows:

\begin{center}
$EST(INV)^j = \sum_{i = 1}^{k} \{\frac{p_i}{\sum_{x = 1}^{k} p_x}$, if $r_{i}^{j} = -1$; \\
\hspace{0.5cm} $ -\frac{1- p_i}{\sum_{x = 1}^{k} (1-p_x)}$, if $r_{i}^{j} = 1$\}.
\end{center}

$EST(INV)^j$  calculates the sum of weighted checking results for all invariants in $INV$ for iteration $j$. The estimation function then calculates the averaged result for the last $w$ consecutive iterations (until $j$):

\begin{center}
$EST(INV)^{j - (w - 1) , j} =  \frac{1}{w}\sum_{i = j - (w - 1)}^{j} EST(INV)^i$.
\end{center}

This averaged value falls in the range of [$-1$, $1$], and a value closer to $1$ would be a strong indicator of a failing execution (i.e., having entered an abnormal state). Like existing work, CoMID needs to set up a threshold for this value to decide whether a monitored execution is failing. Since this value's fluctuation can be largely caused by the uncertainty, we assume that its distribution corresponds to that of the specific uncertainty type a cyber-physical program is experiencing. Then based on the specific uncertainty type (i.e., its error range [$-U$, $U$] and distribution $D$), CoMID sets up the threshold $\Delta$ by solving the uncertainty's $C$-confidence interval equation, i.e., $Pr(x\in[-U \times \Delta, U \times \Delta]) = C$, where $Pr(x)$ is the probability function for distribution $D$). For subjects such as the NAO robot and UAVs in our later evaluation, CoMID sets $w$ = 5 and $C$ = 90\%. The former suggests 2--3 seconds before CoMID makes a decision, which is sufficient for such low-speed subjects to take new actions (customizable by application domains). The latter suggests that CoMID plans to hold a confidence level of 90\% for its made decisions (also customizable by application domains). In the confidence interval equation, the probability function for most uncertainty types follow common models~\cite{toit}, and this facilitates the equation's solution. For example, if a specific certain type follows the uniform distribution, $\Delta$ would be solved to be 0.9; if it follows the normal distribution, $\Delta$ would be 0.65. By doing so, CoMID sets up the threshold $\Delta$ for deciding whether an averaged \emph{EST} value implies the prediction of a failing execution, i.e., by checking whether the value is larger than $\Delta$.

\begin{figure}
\centering \includegraphics[width= 0.5\textwidth]{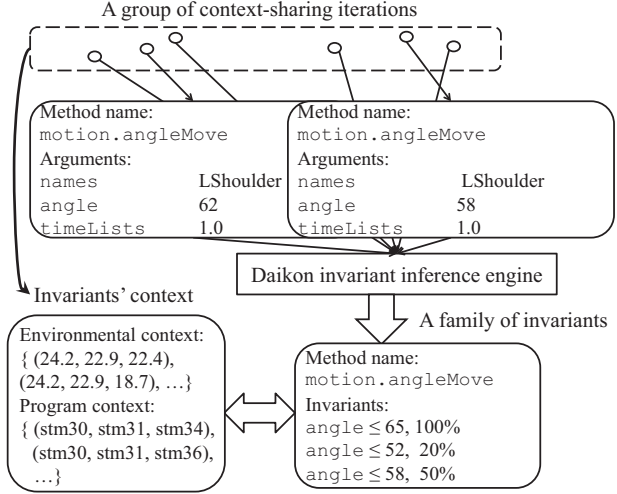}
\caption{Illustration of Step 3: Multi-invariant generation}
\label{fig:step3}
\end{figure}

\begin{figure}
\centering \includegraphics[width= 0.5\textwidth]{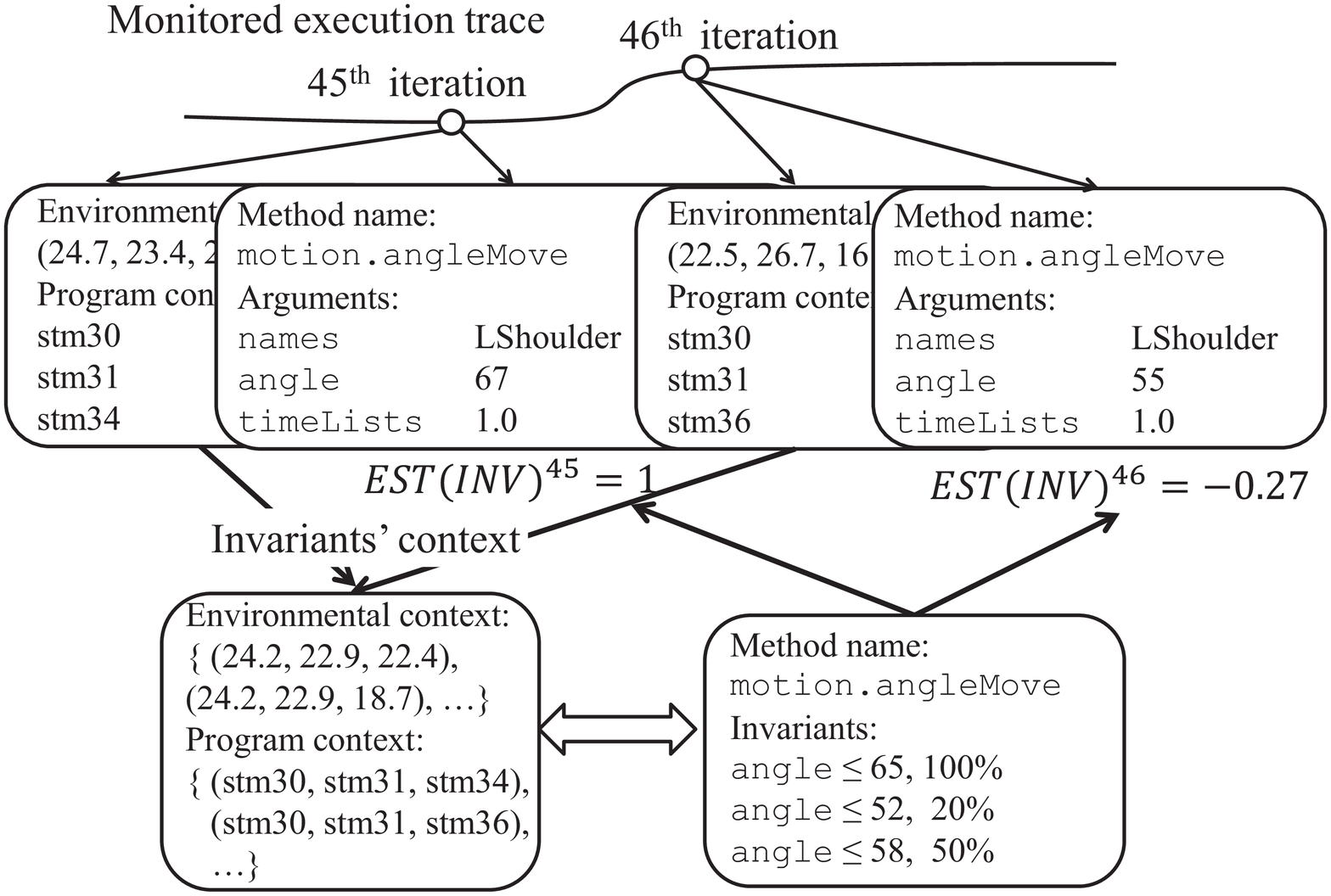}
\caption{Illustration of Step 4: Abnormal-state detection}
\label{fig:step4}
\end{figure}

\textbf{Example.} Consider in our robot example the variable $angle$ for method {\fontfamily{\ttdefault}\selectfont motion.}{\fontfamily{\ttdefault}\selectfont angleMove}{\fontfamily{\ttdefault}\selectfont(names,} {\fontfamily{\ttdefault}\selectfont angle,} {\fontfamily{\ttdefault}\selectfont timeLists)}. Fig.~\ref{fig:step3} illustrates an invariant family for this variable (showing three invariants for example), generated based on one group of context-sharing iterations. In this family, the principal invariant is ``{\fontfamily{\ttdefault}\selectfont angle} $\leq 65, 100\%$'', indicating that the robot's arm should not be raised over $65$ degrees in all cases. This invariant is generated based on all segments (i.e., 100\%) in the concerned group. The other two invariants, namely, ``{\fontfamily{\ttdefault}\selectfont angle} $\leq 52, 20\%$'' and  ``{\fontfamily{\ttdefault}\selectfont angle} $\leq 58, 50\%$'', are generated when only 20\% and 50\% (randomly sampled) segments are used. These invariants' context is also illustrated in Fig.~\ref{fig:step3} (from their corresponding group of segments).

Fig. 6 illustrates how CoMID uses the generated invariant family to detect abnormal states in the runtime monitoring. Consider the 45th and 46th iterations for a monitored execution trace (using two consecutive iterations for example, i.e., $w$ = 2). Suppose that the earlier generated invariant family shares the same context with both iterations. Then CoMID checks all three invariants in the family to decide whether the execution is safe or not. For the 45th iteration, its execution violates all the three invariants, and thus $EST(INV)^{45}$ is calculated to be $1$ ($\frac{1}{1.7} + \frac{0.2}{1.7} +\frac{0.5}{1.7}$). For the 46th iteration, its execution violates only one invariant ``{\fontfamily{\ttdefault}\selectfont angle} $\leq 52, 20\%$'', and thus $EST(INV)^{46}$ is calculated to be $-0.27$ ($-\frac{0}{1.3} + \frac{0.2}{1.7} - \frac{0.5}{1.3}$). So the averaged value of the estimation function for the execution consisting of the 45th and 46th iterations is $0.37$ ($\frac{1 - 0.27}{2})$. If the uncertainty type follows the normal distribution, CoMID would solve the equation to obtain the threshold value to be 0.65, as explained earlier. Then the result (0.37) suggests that the monitored execution is still safe, and that the several invariant violations encountered in these two iterations have been possibly caused by uncertainty.

\section{Evaluation}
\label{sec:exp}
In this section, we evaluate our CoMID approach and compare it with existing work. We select two state-of-the-art invariant generation approaches (i.e., Daikon~\cite{daikon} and ZoomIn~\cite{zoom}) for the comparison. We select three real-world cyber-physical programs, namely, NAO robot, 4-rotor UAV, and 6-rotor UAV, as the experimental subjects. For experimental evaluation, we implement CoMID as a prototype tool in Java 8 and study the following three research questions:

\hangafter 1
\hangindent 1.5em
\textbf{RQ1:} \emph{How does CoMID compare with existing work in detecting abnormal states for cyber-physical programs in terms of effectiveness and efficiency?}

\hangafter 1
\hangindent 1.5em
\textbf{RQ2:} \emph{How does CoMID's configuration (e.g., enabling either or both built-in technique(s) for improving the generated invariants, and setting up which DoS threshold value for distinguishing different program contexts in the invariant generation) affect its effectiveness?}

\hangafter 1
\hangindent 1.5em
\textbf{RQ3:} \emph{How useful is CoMID-based runtime monitoring by invariant generation and checking for cyber-physical programs?}

\subsection{Experimental Preparation and Subjects}
We instrument the three experimental subjects to record their program variable-value and context information during their executions. We use Daikon as the baseline engine for generating invariants from these subjects' execution traces. Besides the invariant templates internally supported by Daikon, we additionally add polygon invariant templates into Daikon, as suggested by existing work~\cite{sebastian,poly} on the runtime monitoring for cyber-physical programs. For fair comparisons, the newly-added invariant templates are used by all the three approaches under the comparison. Note that CoMID is itself independent of the used invariant templates, and this feature makes it general to common cyber-physical programs.

The three experimental subjects are from different companies or universities. The commercial NAO robot program contains 300 LOC (Python-based, with five methods). The two UAV programs are developed by professional electrical engineers, and contain 1,500 LOC (Java-based, with 24 methods) and 4,000 LOC (C-based, with 35 methods), respectively.

\subsection{Experimental Design and Setup}
\textbf{Execution-trace collection.} In the experiments, all invariants should be generated based on the execution traces collected from the selected experimental subjects. For experimental purposes, we design various scenarios for our experimental subjects to run with, and collect their execution traces accordingly. One execution trace is a concrete run of a subject program in a certain scenario. We decide whether an execution trace is \emph{safe} or not (i.e., the \emph{oracle}) according to its corresponding program's behavior and whether its associated failure conditions have been triggered. For the NAO robot (subject \#1), we decide based on its safety (e.g., the robot should never fall into the ground or crash into any obstacle) and liveness (e.g., the robot should not be trapped in a small area). For the UAVs (subjects \#2 and \#3), we decide based on its safety (e.g., a UAV should never fall into the ground or land outside a destination area) and stableness (e.g., a UAV should never lose its height quickly in short time or lose its balance in the air). Besides, if any failure condition is triggered, its corresponding subject program is directly decided to be \emph{unsafe} in its execution. Based on such oracle information (safe or unsafe), we can later judge whether a specific approach under comparison gives a correct prediction or not (i.e., passing vs. safe, and failing vs. unsafe). With the oracle preparation, we test totally six scenarios and collect 1,200 execution traces for the three experimental subjects.

For the NAO robot (subject \#1), we design a $3m \times 3m$ indoor area (including random obstacles and different floor materials) for free exploration. We collect a total of 200 execution traces, including 127 safe ones and 73 unsafe ones. We also build a simulated space with the same settings by the official NAO's emulator Webots~\cite{webot}, and collect 600 execution traces, which include 454 safe ones and 146 unsafe ones. We note that the Webots emulator also supports uncertain environmental sensing internally, and thus its emulated executions are accompanied with uncertainty naturally. However, both the subject program and all the approaches under comparison are unaware of such uncertainty. For ease of presentation, we use \emph{NAO-f} and \emph{NAO-e} to denote the two scenarios, i.e., field setting and emulation setting for the NAO robot, respectively.

For the 4-rotor UAV (subject \#2), we design three field scenarios and collect 100 execution traces for each scenario due to battery constraints. In the first scenario, the UAV takes off from a starting point and lands at a remote destination. We collect 68 safe execution traces and 32 unsafe ones. In the second scenario, the UAV carries some balancing weight during its flying. We collect 71 safe execution traces and 29 unsafe ones. In the last scenario, the UAV conducts extra actions in addition to its normal flying plans, e.g., hovering and turning around. We collect 64 safe execution traces and 36 unsafe ones. Similarly, we use \emph{4-UAV-s1}, \emph{4-UAV-s2}, and \emph{4-UAV-s3} to denote the three scenarios, respectively.

For the 6-rotor UAV (subject \#3), similarly it is scheduled to fly from a starting point to a remote destination. We collect 100 execution traces, including 76 safe executions and 24 unsafe ones. We design one field scenario for experiments and use \emph{6-UAV} to denote this scenario.

\textbf{Experimental procedure.} With collected execution traces from various scenarios, all the approaches under comparison (i.e., CoMID, Daikon, and ZoomIn) generate invariants from these traces, which are evaluated for their qualities, in order to answer our three research questions. All the experiments are conducted on a commodity PC with an Intel(R) Core(TM) i7 CPU @4.2GHz and 32GB RAM. For each scenario, we run CoMID, Daikon, and ZoomIn on safe execution traces to generate invariants, respectively. Then we use safe and unsafe execution traces to validate their generated invariants in detecting abnormal states for the three experimental subjects. We use 10-fold cross-validation in our experiments. More specifically, for each scenario we divide the set of safe execution traces into ten subsets of the same size. One subset of safe execution traces (named the \emph{safe set}) and the set of unsafe execution traces (named the \emph{unsafe set}) are retained for validation. The remaining nine subsets of safe execution traces are used for invariant generation. We repeat this generation and validation process ten times and average their results as the final results for discussion.

To answer research question RQ1 (effectiveness and efficiency), we compare the invariants generated by the three approaches. For each approach, we first study the number of its generated invariants and the percentage of these invariants that can also be generated by other approaches. Since CoMID uses multi-invariant detection, we consider only its \emph{principal invariants} for a fair comparison. We then study the effectiveness and efficiency of the invariants generated by the three approaches in detecting abnormal states for cyber-physical programs. We measure the effectiveness by the \emph{true-positive} rate (TP, i.e., the percentage of unsafe execution traces that are predicted to be failing) for the unsafe set, and by the \emph{false-positive} rate (FP, i.e., the percentage of safe execution traces that are predicted to be failing) for the safe set. Finally, we compare the efficiency for the three approaches by their time costs on invariant generation and checking.

To answer research question RQ2 (impact of configuration), we study CoMID's effectiveness (TP and FP) with its different configurations enabled: (1) on whether to enable one or both built-in technique(s) for improving the generated invariants, i.e., enabling context-based trace grouping only (Context), enabling multi-invariant detection only (Multi), or enabling both techniques (CoMID); (2) on how to set up a DoS threshold value for distinguishing program contexts in the invariant generation, i.e., from 0.6 to 1.0 with a pace of 0.1 (0.8 as the default setting, as explained in Section III.A).

\begin{table*}
\centering
\caption{OVERVIEW OF THE GENERATED INVARIANTS BY THE THREE APPROACHES}
\label{tab1}
\begin{tabular}{|l|r|c|c|r|c|c|r|c|c|}
  \hline
   & \multicolumn{3}{|c|}{CoMID} & \multicolumn{3}{|c|}{Daikon} & \multicolumn{3}{|c|}{ZoomIn}\\
  \cline{2-10}
   & \makecell[c]{Inv} & TP (\%) & FP (\%) & \makecell[c]{Inv} & TP (\%) & FP (\%) & \makecell[c]{Inv} & TP (\%) & FP (\%)\\
  \hline
  NAO-f & 1,157 (33.0\%) & 85.9	& 18.3 & 978 (39.1\%) & 68.6 & 56.0 & 979 (38.0\%) & 78.5 &	43.9\\
  \hline
  NAO-e & 1,313 (32.4\%) & 90.3 & 13.9  & 1,117 (38.1\%) & 79.1 & 44.0 & 1,117 (38.1\%) & 84.6 & 33.9 \\
  \hline
  4-UAV-s1 & 860 (39.0\%) & 95.0 & 15.6 & 577 (58.1\%) & 77.5 & 40.0 & 577 (58.1\%) & 84.3 & 27.2 \\
  \hline
  4-UAV-s2 & 802 (36.3\%) & 93.9 & 7.1  & 570 (51.1\%) & 65.7 & 30.7 & 570 (51.1\%) & 79.1 & 17.2\\
  \hline
  4-UAV-s3 & 933 (30.2\%) & 90.8 & 29.1 & 609 (46.3\%) & 75.5 & 49.4 & 609 (46.3\%) & 80.6 & 35.9 \\
  \hline
  6-UAV & 1,803 (33.0\%) & 92.0 & 12.2  & 1,527 (39.0\%) & 83.4 & 30.7 & 1,527 (39.0\%) & 85.9 & 18.9 \\
  \hline
\end{tabular}
\end{table*}

The first two research questions study the quality of CoMID's generated invariants based on offline execution traces that have been collected in advance. Research question RQ3 aims to study CoMID's usefulness in the runtime monitoring, i.e., investigating how its generated invariants behave when monitoring the three experimental subjects' online executions. Without CoMID-based runtime monitoring, the three experimental subjects can rely on only their built-in \emph{protection mechanisms} when their corresponding failure conditions are triggered. For example, when the robot is falling into the ground, it would control to stop walking and crouch on its knees; when a UAV is falling into the ground, it would control to stop rotating its wings. Such protection mechanisms can prevent the robot and UAVs from being damaged by the failures, but their planned tasks already fail. With CoMID-based runtime monitoring, the three experimental subjects can take \emph{remedy mechanisms} in advance once CoMID detects abnormal states (i.e., predicting the current execution to be failing). For example, the robot would stop walking, stand for two seconds, and then walk toward a different direction; a UAV would stop landing, reinitiate the flying plan, and then seek to land after two seconds. Although such remedy mechanisms can delay the subjects' planned tasks, they should be able to help avoid upcoming failures that would otherwise occur if no remedy mechanism is taken. We study CoMID's usefulness by comparing the difference between using and not using CoMID-based runtime monitoring.

To answer RQ3 (usefulness), we study how CoMID-based runtime monitoring helps the three experimental subjects on preventing their failures. The failure data without CoMID-based runtime monitoring can be obtained from earlier collected execution traces for the three experimental subjects in answering RQ1 and RQ2. For obtaining the failure data with CoMID-based runtime monitoring, we run the three experimental subjects enabled with CoMID-based runtime monitoring and remedy mechanisms 100 times for each scenario, and average their results. Then we can calculate and compare the success rates for the three experimental subjects from the failure data. In addition, since the remedy mechanisms can delay the subjects' planned tasks, we study their impact by measuring and comparing the subjects' task-completion time (i.e., when a robot finishes its exploration task, and a UAV finishes its flying and landing tasks) for those non-failure executions.

\subsection{Experimental Results and Analyses}

\textbf{RQ1 (effectiveness and efficiency).} Table~\ref{tab1} gives an overview of our experimental results on the quality of the generated invariants by the three approaches under comparison. It includes the number of generated invariants (Inv), true positive rate (TP) in detecting abnormal states for the unsafe set, and false positive rate (FP) in detecting abnormal states for the safe set. The percentage data in brackets after the invariant numbers give the proportions of the concerned invariants that can also be generated by other approaches. In general, CoMID generates more invariants than Daikon and ZoomIn (17.5--53.2\% more, for different scenarios), even if we consider its principal invariants only. This is because CoMID generates different invariants to govern the program behavior for different situations by distinguishing different program and environmental contexts. Daikon and ZoomIn generate the same numbers of invariants since ZoomIn internally uses Daikon as its invariant inference engine, although they check these invariants in different ways in the runtime monitoring, as we show later.

Besides, we observe that the invariants generated by CoMID are quite different from those generated by the other two approaches. For example, 30.2--39.0\% of CoMID's invariants can be generated by the other two approaches, but 38.1--58.1\% of the other two approaches' invariants can also be generated by CoMID. Considering that the number of CoMID's generated invariants is larger than those of the other two approaches' generated invariants, this suggests that CoMID generates much more invariants that are unique from those generated by the other two approaches.

It is important to know whether these unique invariants bring the positive or negative impact on detecting abnormal states for the three experimental subjects. We observe from the Table I that these unique invariants enable CoMID to achieve a higher TP and a lower FP. For example, CoMID's TP is 8.6--28.2\% higher than Daikon and 5.7--14.7\% higher than ZoomIn, and at the same time, CoMID's FP is 18.6--37.6\% lower than Daikon and 6.8--25.5\% lower than ZoomIn. A high TP implies the ability of capturing various cases of abnormal states, and at the same time, a low FP implies that this ability is not achieved by the cost of overfitting the generated invariants to specific cases. Therefore, this result suggests that CoMID's generated invariants are of a high quality, by achieving both a high TP and a low FP. It also indicates that CoMID deserves its efforts on particularly addressing the iterative execution and uncertain interaction characteristics of cyber-physical programs. For the iterative execution, ZoomIn partially uses program contexts to distinguish different scopes for different invariants, and thus performs better than Daikon, which does not consider any context at all. For the uncertain interaction, different levels of uncertainty result in CoMID's varying leading advantages in FP for different experimental subjects. For example, compared with ZoomIn, CoMID achieves a 20.0--25.5\% lower FP for the NAO robot, and a 6.8--11.2\% lower FP for the two UAVs. The differences are due to the fact that the NAO robot suffers more uncertainty (e.g., installed more types of sensors) than the two UAVs.

We then compare the efficiency for the three approaches in generating invariants and checking these invariants for detecting abnormal states. Fig. 8-a compares these approaches' time costs in generating invariants. We observe that CoMID spends 18.7--43.6\% more time than Daikon and 8.9--23.5\% more than ZoomIn in generating invariants. Daikon spends the least time due to its straightforward strategy of invariant generation by overlooking all contexts. CoMID's higher time cost is due to its constituent techniques of context-based trace grouping and multi-invariant generation for improving the quality of generated invariants. For the former, CoMID groups context-sharing iterations to make its generated invariants fitter to specific program behaviors, bringing up its TP in detecting abnormal states. For the latter, CoMID uses multiple invariants to alleviate the impact of uncertainty, bringing down its FP in detecting abnormal states.

\begin{figure}
  \centering
  \subfigure[For generating invariants]{
  \includegraphics[width= 0.45\textwidth]{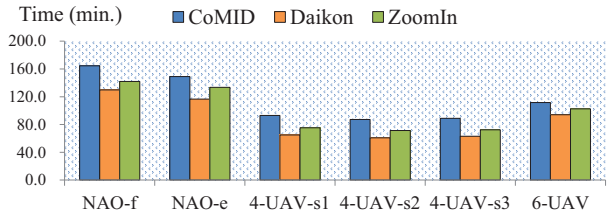}}
  \label{fig:gen}
  \subfigure[For checking invariants]{
  \includegraphics[width= 0.45\textwidth]{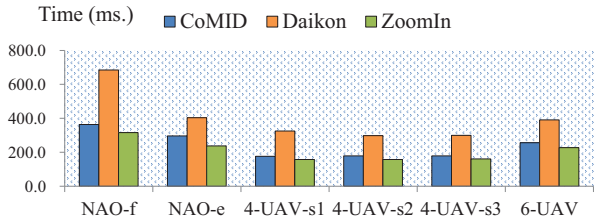}}
  \label{fig:dect}
  \caption{Efficiency comparison for CoMID, Daikon, and ZoomIn}
\end{figure}

Fig. 8-b compares the three approaches' time costs in checking invariants for runtime monitoring. We observe that CoMID spends 36.3--88.5\% less time than Daikon. Although CoMID uses multiple invariants to decide abnormal states, its context-based trace grouping technique enables it to focus on much fewer invariants specific for each iteration a cyber-physical program encounters. Daikon, instead, has to check each invariant in each iteration, resulting in its high time cost in detecting abnormal states. Regarding ZoomIn, it uses only statement-coverage information to select invariants for checking in each iteration, resulting in its lowest time cost (no other overhead), but still comparable to CoMID's time cost, as shown in Fig. 8-b.

Therefore, we answer\textbf{ research question RQ1} as follows.
\begin{center}
\fbox{\begin{minipage}{0.95\linewidth}
\emph{CoMID generates and checks invariants to detect abnormal states for cyber-physical programs effectively and efficiently. It achieves a higher TP (5.7--28.2\% higher) and a lower FP (6.8--37.6\% lower) than Daikon and ZoomIn. Although it spends more time in generating invariants (offline), its invariant checking (online) is comparably efficient as ZoomIn and much more efficient than Daikon.}
\end{minipage}}
\end{center}

\begin{figure}
  \centering
  \subfigure[TP comparison]{
  \includegraphics[width= 0.45\textwidth]{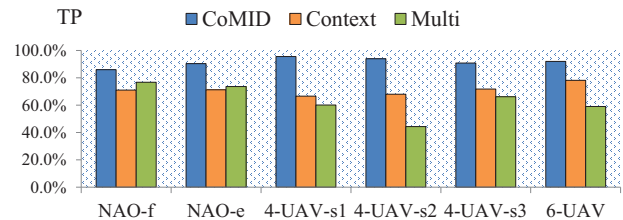}}
  \subfigure[FP comparison]{
  \includegraphics[width= 0.45\textwidth]{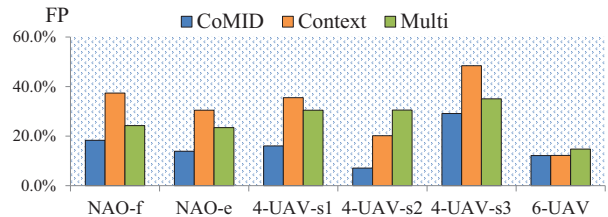}}
  \caption{Effectiveness comparison for CoMID, Context, and Multi}
  \label{fig:feature}
\end{figure}

\textbf{RQ2 (impact of configuration)}. We study the impact of configurations on CoMID's effectiveness from two lines. First, CoMID can be configured with its two built-in techniques (context-based trace grouping and multi-invariant detection) individually enabled. Fig.~\ref{fig:feature} compares the effectiveness in terms of TP and FP for the original CoMID (CoMID), CoMID with only context-based trace grouping enabled (Context), and CoMID with only multi-invariant detection enabled (Multi). We observe that when detecting abnormal states for the unsafe set, Context performs more effectively than Multi in four UAV scenarios (5.7--23.6\% higher TP), while Multi performs more effectively than Context in two NAO scenarios (2.3--5.7\% higher TP). As analyzed earlier, the NAO robot suffers more uncertainty than the two UAVs due to its complicated sensing and physical behavior, and thus Multi helps more than Context for the two NAO scenarios on suppressing the impact of uncertainty. For the four UAV-related scenarios, their uncertainty is relatively lighter, and thus Context exhibits more significant advantages. Of course, when combining the two techniques together, CoMID always produces the best results (9.2--49.5\% higher TP). On the other hand, when suppressing false alarms for the safe set, Multi performs more effectively than Context in four scenarios (5.0--13.3\% lower FP), and Context performs more effectively than Multi in the other two scenarios (2.5--10.4\% lower FP). The differences are mainly caused by different levels of uncertainty in the respective scenarios. Still, CoMID again produces the best results (0.1--23.5\% lower FP). Considering that Context and Multi behave better in different scenarios (complementing each other) and CoMID always produces the best results, CoMID's two techniques (context-based trace grouping and multi-invariant detection) are both useful for improving its effectiveness by achieving a high TP and a low FP.

\begin{figure}
  \centering
  \subfigure[TP comparison]{
  \includegraphics[width= 0.45\textwidth]{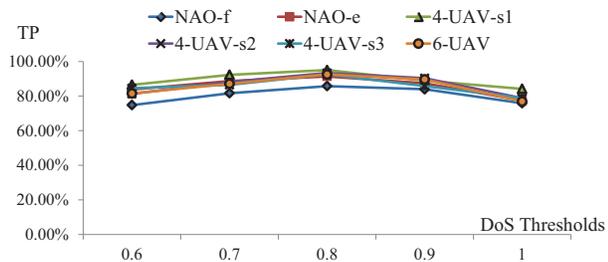}}
  \subfigure[FP comparison]{
  \includegraphics[width= 0.45\textwidth]{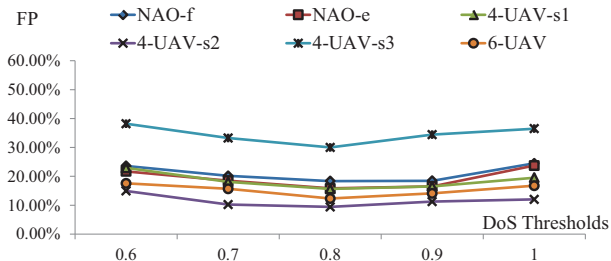}}
  \caption{Effectiveness comparison for CoMID with different DoS threshold values}
  \label{fig:dos}
\end{figure}

Second, CoMID can also be configured to use different DoS threshold values for distinguishing different program contexts in generating invariants. As mentioned earlier, CoMID uses a default DoS threshold value of 0.8 as suggested by existing work~\cite{zoom}, and here we study the impact of this value choice (from 0.6 to 1.0 with a pace of 0.1) on CoMID's effectiveness. Fig.~\ref{fig:dos} compares CoMID's effectiveness in terms of TP and FP with different DoS threshold values. We observe that in all six scenarios, CoMID with the value of 0.8 indeed behaves the best in both TP and FP. Nevertheless, the winning extents are not that large, and the extent on TP (1.8--15.6\% higher) is a bit more than that on FP (0.1--7.9\% lower). In addition, we observe that the impact of different DoS threshold values varies across different scenarios. For example, in scenario NAO-f, the TP for threshold 0.9 behaves slightly better than that for threshold 0.7, while in scenario NAO-e, the latter behaves slightly better than the former. This result suggests that CoMID's effectiveness might be further improved if its DoS threshold value can be tuned adaptively for specific cyber-physical programs. Currently, we make CoMID take the default value of 0.8 for simplicity, and we leave its adaptive tuning to future work.

Therefore, we answer \textbf{research question RQ2} as follows.
\begin{center}
\fbox{\begin{minipage}{0.95\linewidth}
\emph{CoMID's configurations affect its effectiveness. First, CoMID's two built-in techniques are both useful. When the uncertainty affecting the experimental subjects is relatively light, CoMID with only context-based trace grouping enabled already behaves quite well. When the uncertainty is relatively heavy, CoMID with only multi-invariant detection enabled behaves better. In either way, combing both techniques (i.e., a fully-fledged CoMID) produces the best results. Second, CoMID's setting of its DoS threshold value for distinguishing different program contexts also affects its effectiveness, but not significantly. Its current default value of 0.8 already makes it work satisfactorily for the experimental subjects.}
\end{minipage}}
\end{center}

\begin{figure}
\centering \includegraphics[width= 0.5\textwidth]{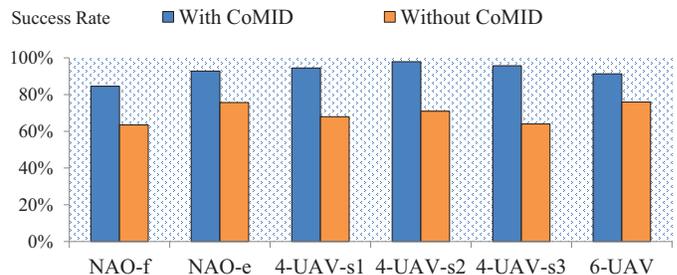}
\caption{Success rate for monitored cyber-physical programs }
  \label{fig:suc}
\end{figure}

\begin{figure}
\centering \includegraphics[width= 0.5\textwidth]{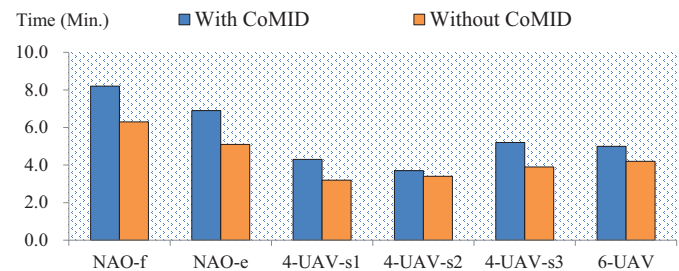}
\caption{Average task-completion time for monitored cyber-physical programs }
  \label{fig:time}
\end{figure}

\textbf{RQ3 (usefulness)}. Finally, we study how CoMID-based runtime monitoring helps the three experimental subjects on preventing their potential failures. Fig.~\ref{fig:suc} compares the success rate for the three experimental subjects in the six scenarios, based on their failure data with (``with CoMID'') and without (``without CoMID'') CoMID-based runtime monitoring. We observe that CoMID indeed helps improve the success rate by 15.3--31.7\% (avg. 23.1\%) across different scenarios. This result echoes our earlier experimental results on CoMID's high TP and low FP performance. In addition, as mentioned earlier, the CoMID-based runtime monitoring and remedy mechanisms can delay the three experimental subjects' planned tasks, thus trading for higher safety (i.e., fewer failures). So we study such  impact. Fig.~\ref{fig:time} compares the average task-completion time for non-failure executions of the three experimental subjects with (``with CoMID'') and without (``without CoMID'') CoMID-based runtime monitoring. We observe that CoMID indeed increases the subjects' task-completion time by 8.8--35.2\% (avg. 26.8\%). We consider such slowdown extent acceptable for subjects that require high safety assurance. In fact, the delay is largely due to the safety control before reinitializing the tasks (e.g., a robot stands for two seconds and then restarts walking, and a UAV restarts to land after two seconds), which is customizable by different application domains.


Therefore, we answer \textbf{research question RQ3} as follows.
\begin{center}
\fbox{\begin{minipage}{0.95\linewidth}
\emph{CoMID's capability of generating and checking invariants for runtime monitoring can effectively prevent the experimental subjects from entering potential failures. CoMID helps improve the subjects' success rate in their task executions by 15.3--31.7\%, with a cost of 8.8--35.2\% longer task-completion time.}
\end{minipage}}
\end{center}
\subsection{Threats to Validity}
One major concern on the validity of our experimental conclusions is the selection of experimental subjects in our evaluation. We select only three experimental subjects, which may not allow our conclusions to be generalized to more other subjects. Nevertheless, a comprehensive evaluation requires the support of suitable environments for experiments, which should be both observable and controllable. This requirement restricts our choice of possible experimental subjects. To alleviate this threat, we try to make our subjects \emph{realistic} by selecting real-world cyber-physical programs. Besides, we make the subjects \emph{diverse} by requesting them to cover different functionalities (e.g., automated area exploration, planned flying, and smart obstacle avoidance), and to run on different platforms (e.g., Python-based NAO robot, Java-based UAV, and C-based UAV). By doing so, we try to alleviate as much as possible potential threat to the external validity of our experimental conclusions. Still, evaluating CoMID on more comprehensive cyber-physical programs and platforms deserves further efforts.

Another concern is about relating the detection of an abnormal state to an execution's failure result; such factor may pose threat to internal validity of our experimental conclusions on an approach's TP and FP performance. The reason is that when an abnormal state is detected by an approach, one seems not able to clearly relate the detection to the current execution's upcoming failure, considering that their time interval can vary. To address this problem, we particularly design to measure TP for unsafe executions and FP for safe executions only: (1) for an unsafe execution, if an approach never detects any abnormal state, such result suggests its weakness (it should detect), and so we choose to check whether the approach reports the detection of any abnormal state, i.e., TP; (2) on the other hand, for a safe execution, if an approach reports the detection of any abnormal state, such result also suggests its weakness (it should not detect), and so we directly check whether the approach produces such false alarms, i.e., FP. In addition, to further alleviate the potential threat, we additionally study in research question RQ3 whether CoMID-based runtime monitoring indeed helps prevent the experimental subjects  from entering failures, i.e., by measuring and comparing their success rates in task executions. All together, we strive our best efforts to evaluate CoMID's experimental and practical usefulness for cyber-physical programs.

\section{Related Work}
\label{sec:rela}
In this section, we discuss representative related work in recent years, on testing cyber-physical programs, generating program invariants, and runtime monitoring, respectively.

\textbf{Testing cyber-physical programs.} Cyber-physical programs are featured with context-awareness, adaptability, and uncertain program-environmental interactions, which bring substantial challenges to their quality assurance. To address this problem, various techniques have been proposed for effective testing of such programs. For example, Fredericks et al.~\cite{seams14} used utility functions to guide the design and evolution of test cases for cyber-physical programs. Xu et al.~\cite{xu@jss12} proposed monitoring common error patterns at the runtime of cyber-physical programs, to identify defects in their adaptation logics when interacting with uncertain environments. Ramires et al.~\cite{ramires} explored specific combinations of environmental conditions to trigger specification-violating behaviors in adaptive systems. Yi et al.~\cite{sit} presented a white-box sampling-based approach to systematically exploring the state space of an adaptive program, by filtering out unnecessary space samplings whose explorations would not contribute to detecting program faults. These pieces of work exploited different observations to strengthen their testing effectiveness, but relied mostly on human-written or domain-specific properties for defining abnormal or error states in executing programs. Our CoMID approach complements such work by assisting their fault-detection capabilities from checking trivial failure conditions (e.g., system crashes) to comprehensive errors (e.g., various types of error state) with automatically generated invariants.

\textbf{Generating program invariants.} Invariants play an important role in program analysis and runtime monitoring towards the quality assurance for programs with complex logics. As a representative approach of automated invariant generation, Daikon~\cite{daikon} inferred pre-conditions and post-conditions for methods executed in a program, by collecting program execution information and using pre-defined templates to derive invariants from the collected information. DySy~\cite{dysy} followed a similar way and used branch conditions to derive more types of invariants. Eclat~\cite{elcat} took automated invariant generation one step further, by learning a model from assumed executions and identifying inputs that do not match the learned model. Jiang et al.~\cite{sebastian} derived invariants by observing messages exchanged between system nodes, and specified operational attributes for robotic systems based on these messages. Zhang et al.~\cite{idis} used symbolic execution as a feedback mechanism to refine the set of candidate invariants generated by Daikon. Carzaniga et al.~\cite{car} proposed cross-checking invariant-alike oracles by exploiting intrinsic redundancy of software systems. Different from these pieces of existing work, our CoMID approach additionally considers the impact of contexts on invariant generation (to restrict invariants' effective scopes) and that of uncertainty on invariant checking (to suppress false alarms), specially catered for the characteristics of cyber-physical programs.

\textbf{Runtime monitoring.} By means of invariant checking, one is able to detect abnormal states or anomalous behaviors in a program's execution and take remedy actions if necessary, thus helping improve the program's quality at runtime. For example, Zheng et al.~\cite{zhang} mined predicate rules that specify what must hold at certain program points (e.g., branches and exit points) for runtime monitoring. Raz et al.~\cite{Raz} derived constraints on values returned by data sources, and identified abnormal values based on the derived constraints. Pastore et al.~\cite{zoom} used the statement-coverage information in a program's execution to improve the precision of abnormality detection. Nadi et al.~\cite{nadi} extracted configuration constraints from program code, and used the constraints to enforce expected runtime behaviors. Xu et al.~\cite{dsn16} collected the calling contexts of method invocations, and used the contexts to distinguish a program's different behaviors under different scenarios. These pieces of work shared a common assumption that a program execution's anomalous behaviors can be discovered by checking newly collected execution data against earlier derived constraints from assumed normal executions. While this assumption is generally correct, cyber-physical programs' two characteristics, i.e., iterative execution and uncertain interaction as discussed earlier, make these pieces of work less effective. The main reason is that different iterations in a cyber-physical program's execution can face different situations and undertake different strategies to handle these situations. Then a straightforward invariant-checking approach can easily generate false alarms when the derived invariants' scopes differ and the impact of uncertainty is overlooked. Our CoMID approach specifically addresses this problem and thus complements existing work on effective runtime monitoring.

\section{Conclusion and Future work}
\label{sec:end}
In this article, we present a novel approach, CoMID, for effectively generating and checking invariants to detect abnormal states for cyber-physical programs. CoMID distinguishes different contexts for invariants and makes them context-aware, so that its generated invariants can be effective for varying situations and at the same time robust to uncontrollable uncertainty faced by cyber-physical programs. Our experimental evaluation with real-world cyber-physical programs validates CoMID's effectiveness in improving the true-positive rate and reducing the false-positive rate in detecting abnormal states, as compared with two state-of-the-art invariant generation approaches.

CoMID still has room for improvement. For example, it currently relies on pre-decided models about the uncertainty distribution a cyber-physical program experiences in its multi-invariant detection. We plan to relax this requirement and explore dynamic calibration techniques to refine such uncertainty models in order to support more application scenarios in future. Besides, CoMID currently uses the default DoS threshold value of 0.8 as suggested by existing work~\cite{zoom}. In experiments, we observe the opportunities in which different threshold values can bring higher runtime monitoring qualities for different scenarios. Therefore, it is also worth exploring how to design adaptive DoS threshold tuning for further refined invariant generation and checking. We are working along these lines.
\section*{Acknowledgment}
The authors wish to thank the editor and anonymous reviewers for their valuable comments on improving this
article. This work was supported in part by National Key R\&D Program (Grant \#2017YFB1001801) and National Natural Science Foundation (Grants \#61690204, \#61472174) of China. The authors would also like to thank the support of the Collaborative Innovation Center of Novel Software Technology and Industrialization, Jiangsu, China. This work was supported in part by National Science
Foundation under grants no. CCF-1409423, CNS-1513939,
CNS1564274, and the GEM fellowship.

\begin{IEEEbiographynophoto}{Yi Qin}
Yi Qin received his bachelor degree in computer science and technology from Nanjing University, China. He is currently working toward the doctoral degree at the department of computer science and technology at Nanjing University, China. His research interests include software testing and adaptive software systems.
\end{IEEEbiographynophoto}

\begin{IEEEbiographynophoto}{Tao Xie}
Tao Xie received his doctoral degree in computer science from the University of Washington at Seattle, USA. He is a Professor and Willett Faculty Scholar in the Department of Computer Science at the University of Illinois at Urbana-Champaign, USA.  He was the ISSTA 2015 PC Chair and will be an ICSE 2021 PC Co-Chair. He has been an Associate Editor of the IEEE Transactions on Software Engineering (TSE) and the ACM Transactions on Internet Technology (TOIT), along with an Editorial Board Member of Communications of ACM (CACM). His research interests are in software engineering, with recent focus on intelligent software engineering. He is an ACM Distinguished Scientist and an IEEE Fellow.
\end{IEEEbiographynophoto}

\begin{IEEEbiographynophoto}{Chang Xu}
Chang Xu received his doctoral degree in computer science and engineering from The Hong Kong University of Science and Technology, Hong Kong, China. He is a full professor with the State Key Laboratory for Novel Software Technology and Department of Computer Science and Technology at Nanjing University. He participates
actively in program and organizing committees of major international software engineering conferences. He co-chaired the MIDDLEWARE 2013 Doctoral Symposium, FSE 2014 SEES Symposium, and COMPSAC 2017 SETA Symposium. His research interests include big data software engineering, intelligent software testing and analysis, and adaptive and autonomous software systems. He is a senior member of the IEEE and member of the ACM.
\end{IEEEbiographynophoto}

\begin{IEEEbiographynophoto}{Angello Astorga} received his bachelor degree in Computer Science and Engineering with Magna Cum Laude Honors from the Ohio State University. He is currently working toward the doctoral degree at the Department of Computer Science at the University of Illinois at Urbana-Champaign, USA. His research interests include software engineering and software verification.
\end{IEEEbiographynophoto}

\begin{IEEEbiographynophoto}{Jian Lu}
Jian Lu received his doctoral degree in computer science and technology from Nanjing University, China. He is a full professor with the Department of Computer Science and Technology and Director of the State Key Laboratory for Novel Software Technology at Nanjing University. He has served as a Vice Chairman of
the China Computer Federation since 2011. His research interests include software methodologies, automated software engineering, software agents, and middleware systems.
\end{IEEEbiographynophoto}
\end{document}